\documentclass{IOS-Book-Article}

\usepackage{mathptmx}
\usepackage{soul}\setuldepth{article}
\usepackage{booktabs}
\usepackage{graphicx}
\usepackage{subcaption}
\usepackage{tabularx}

\usepackage{array}  
\usepackage{booktabs} 
\usepackage{adjustbox} 
\usepackage{multirow} 

\def\hb{\hbox to 11.5 cm{}}

\begin{document}

\pagestyle{headings}
\def\thepage{}
\begin{frontmatter}              

\title{Getting in the Door: Streamlining Intake in Civil Legal Services with Large Language Models }

\markboth{}{September 2024\hb}

\author[A]{\fnms{Quinten} \snm{Steenhuis}\orcid{0009-0001-0110-064X}}%
and
\author[B]{\fnms{Hannes} \snm{Westermann}\orcid{0000-0002-4527-7316}}\footnote{And special thanks to additional developer, Toby Fey}

\address[A]{Co-director, Suffolk University Law School Legal Innovation and Technology Lab}
\address[B]{Assistant Professor, Maastricht University Law and Tech Lab}

\begin{abstract}
Legal intake, the process of finding out if an applicant is eligible for help from a free legal aid program, takes significant time and resources. In part this is because eligibility criteria are nuanced, open-textured, and require frequent revision as grants start and end. In this paper, we investigate the use of large language models (LLMs) to reduce this burden. We describe a digital intake platform that combines logical rules with LLMs to offer eligibility recommendations, and we evaluate the ability of 8 different LLMs to perform this task. We find promising results for this approach to help close the access to justice gap, with the best model reaching an F1 score of .82, while minimizing false negatives.
\end{abstract}

\begin{keyword}
large language models\sep access to justice\sep
civil legal aid\sep law\sep machine learning \sep natural language processing \sep legal triage \sep housing intake
\end{keyword}
\end{frontmatter}
\markboth{December 2024\hb}{December 2024\hb}

\section{Introduction}
Around the world, accessing meaningful help for civil legal problems means hitting closed door after closed door. In the U.S., the first door—reaching a nonprofit legal aid lawyer—usually only opens after a poor person waits in a telephone intake queue for hours.

The intake screening challenge is an important piece of the ``access to justice" gap, which affects up to 92\% of the poor worldwide.\cite{legal_services_corporation_justice_nodate, world_justice_project_global_2019, currie2009legal, savage2022experiences, Westermann_thesis_2023}, Unresolved legal issues, like housing, custody, guardianship and domestic violence claims, can harm individuals and burden society. \cite{semple2015cost, savage2022experiences, world_justice_project_global_2019, farrow2016everyday} Legal aid programs, an important part of the solution, have funding to offer assistance only to a limited number of even the eligible applicants.\cite{smith1919justice, houseman2018securing, bba_investing_justice}

To be eligible for legal help, individuals needs to comply with certain \textit{formal requirements}, such as household income and citizenship requirements. The individual also needs to comply with a set of \textit{substantive requirements}, relating to the specific legal problem they face. Because of high demand, an applicant who calls for intake screening may wait on hold for several hours.

The substantive requirements pose a special challenge for automation. \textit{First}, they are in constant flux. As grants start and end and staffing rises and falls, intake priorities can change from week to week. \textit{Second}, the substantive requirements can be highly varied and specific to a single organization. \textit{Third}, determining eligibility for the substantive requirements often relies on nuanced assessments of open-textured terms (such as e.g., ``severity'' of the need for housing repairs).

Online tools have long been seen as valuable for the intake process, offering 24/7 availability without requiring staff time, but while formal requirements can be encoded with logical rules (c.f. \cite{Atkinson2024}) the nuanced and constantly changing substantive requirements are harder to handle with traditional methods.

Our study investigates this legal aid intake challenge and describes a solution for housing intake piloted in the U.S. state of Missouri, in collaboration with four legal aid programs with highly heterogeneous intake requirements.\footnote{Special thanks to Legal Services of Eastern Missouri, Legal Assistance of Western Missouri, Legal Services of Southern Missouri, and Mid-Missouri Legal Services for their assistance in this study.}\footnote{This work was funded in part by the U.S. Department of Housing and Urban Development (HUD).} The pilot employs both hand-coded rules and a large language model to help inform applicants about their likelihood of qualifying for help \textbf{before} they wait in a telephone queue. The platform is described below in section \ref{sec:proposed_framework}. While we piloted the project in Missouri, the diversity of the rules we tested suggest that the results of this study are broadly relevant for legal intake.

\subsection{Research Questions}
In order to understand how LLMs can help with intake of legal aid applicants, we will investigate the following research questions:

\begin{itemize}
\item RQ1 - How accurately can large language models apply intake rules to legal scenarios described by laypersons?
\item RQ2 - What types of errors do LLMs make, and how do they compare to human intake staff? 
\item RQ3 - Can LLMs elicit missing information through high-quality follow-up interactions?
\end{itemize}

\section{Prior Work}
This work builds on a number of important prior contributions.

\subsection{AI for access to justice}

Using AI for access to justice is a growing field with applications in legal information\cite{branting2001advisory, paquin1991loge, westermann_justicebot_2023}, form-filling\cite{steenhuis_beyond_2023, westermann2024dallma}, and dispute resolution \cite{de2021making, schmitz2021intelligent, steenhuis2024ai,Westermann_thesis_2023, steenhuis_making_2019, thompson2015creating, salter2017online, bickel2015online, Branting2022-BRAACM-5, westermann2023llmediatorgpt4assistedonline, zeleznikow2016can, lodder2005developing}.

Prior work in the area of eligibility determinations includes the GetAid system \cite{zeleznikow2002using}, which uses decision trees and Toulmin argument structures to help lawyers determining legal aid eligibility. Atkinson et. al. \cite{Atkinson2024} apply formal methods to make eligibility recommendations for asylum claims in the  European Court of Human Rights. Here, we investigate whether LLMs can determine eligibility upon reading verbatim intake guidelines, which would be easier and faster to deploy than these models.

\subsection{LLMs for Legal Reasoning}
Recent years have shown a wide range of legal reasoning capabilities for LLMs, including passing the bar exam\cite{Katz2024GPT4}, drafting contracts\cite{lam2023applying, wang_llm_contracts}, decision making\cite{dawson2024algorithmic, crootof2023humans, gutierrez2024critical}, annotating legal documents \cite{savelka2023GPT4,savelka2023unlocking}, explaining legal concepts \cite{savelka2023concepts}, performing statutory reasoning \cite{blair2023can,nguyen2023well} and providing legal information \cite{tan2023, cui2024chatlawmultiagentcollaborativelegal}. The LegalBench study measures 20 models on 161 different legal tasks
\cite{guha2023legalbenchcollaborativelybuiltbenchmark}. \cite{prakken2024evaluating} discusses the LLMs ability to perform legal reasoning. Here, we investigate whether they can apply complex legal intake rules (provided as text) to factual scenarios to determine legal intake eligibility.


\subsection{Spotting legal issues and eliciting facts}
Previous work has used machine learning to spot legal issues and direct users to decision support tools.\cite{About_Spot, westermann2023bridging} Our approach differs by relying on zero-shot LLMs to assess eligibility based on textual rules. 

Sometimes, an initial user scenario description may not be sufficient to determine rule applicability. A number of projects have explored the use of AI and LLMs to elicit facts from users. \cite{branting_narrative-driven_2023} uses schemas induced from cases to identify key facts for elicitation. \cite{westermann2024dallma} uses LLMs and logical rules to perform step-by-step assessment of whether relevant facts are present in user descriptions. \cite{goodson_intention_2023} use LLMs to ask follow-up questions in legal aid advice scenarios in order to maximize the understanding of the clients intention and situation. Here, we use LLMs to elicit information relevant to determine the eligibility of intake rules, and evaluate the capability with intake workers.

\section{Proposed framework}
\label{sec:proposed_framework}
We used the free and open source Docassemble framework to build the user-facing housing intake application. \cite{jonathan_pyle_docassemble_2021, lauritsen_substantive_2019, steenhuis_making_2019} The intake application can be accessed on a tenant's mobile phone and is both embedded in a legal help website (MOTenantHelp.org) and referred to in the on-hold message for callers to the phone intake system.

A typical interaction with the application is available on a smart phone, takes about 5 minutes and asks questions on 5 separate screens.

\subsection{Formal eligibility criteria}
We used formal rules, encoded in Python, to determine eligibility based on location and statutory requirements of 42 U.S.C. 2996g(e).

\subsection{Program-specific substantive criteria}\label{sec:intake_rules}
Program-specific rules were stored as plain text inside the Docassemble application for later retrieval. The rules, which we left unmodified, showed diversity in length, formatting, and substance. For example: one program uniquely accepted security deposit disputes, and two others prioritized discrimination cases. Housing subsidies (income-based government support), a common priority, were described using varying key words across the programs. A fourth program, which we did not study, accepted ``all landlord tenant disputes."

Although the platform handles all four programs, we focused on three in our study:
\begin{enumerate}
    \item Eastern Missouri (St. Louis). Intake rules: 3,792 characters.
    \item Mid-Missouri. Intake rules: 1,437 characters.
    \item Western Missouri - Central office (Kansas City). Intake rules: 1,030 characters.
\end{enumerate}

These intake rules are quite complex. Mid-Missouri, for example, has 12 separate grounds for accepting a housing case, each of which can require multiple facts for assessment.

\subsection{Screening with LLM assistance}
\begin{figure}
    \centering
    \includegraphics[width=.8\linewidth]{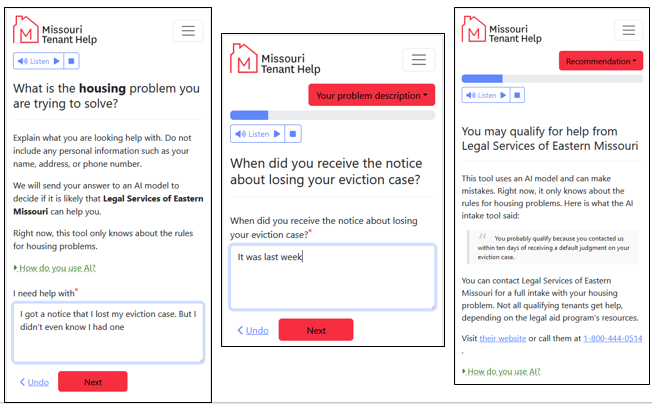}
    \caption{Figure showing progress from problem description, to follow-up question, to a final recommendation to the applicant. The recommendation is qualified (e.g., it says ``you probably qualify") and has a button to allow the applicant to learn more about how AI is used in the decision.}
    \label{docassemble_interview}
\end{figure}
We ask the applicant to describe their legal problem in their own words, without any personally identifiable information. The full set of rules and the problem description are then provided in a prompt to GPT-4-turbo with the temperature set to 0. The LLM is instructed to respond with one of three statuses: qualifies if the applicant appears to meet the criteria, does not qualify if they clearly do not, or a follow-up question if more information is needed. Up to 10 follow-up questions are asked. 

Finally, the applicant is told the LLM's determination of their eligibility. We add a disclaimer that the AI tool can get the decision wrong, and we provide a link to the program's website and phone number to complete a full intake.

\section{Experimental design}

\subsection{Dataset}
To test the LLM-based intake process, we developed two datasets, D1 and D2.

\subsubsection{D1 - Initial Response Dataset}\label{sec:scenarios}

\begin{table}[htbp]
\centering
\caption{Table showing the number of times a certain target outcome (Accept, Deny, Question) is present in the annotated dataset, per jurisdiction and in total.}
\begin{tabular}{|l|rrr|r|c|}
\hline
 &  Accept &  Deny &  Question &  Total  \\

\hline
Eastern Missouri  &       8 &     7 &         1 &     16  \\
Mid-Missouri      &       8 &     6 &         2 &     16  \\
Western Missouri  &       2 &    12 &         2 &     16  \\
\hline
Total             &      18 &    25 &         5 &     48  \\
\hline
\end{tabular}
\label{table_d1}
\end{table}

\begin{table}[h]
\centering
\begin{adjustbox}{max width=\textwidth} 
\begin{tabular}{|p{5cm}|p{5cm}|p{1cm}|} 
\hline
 \textbf{Scenario} & \textbf{Relevant excerpt from rule} & \textbf{Decision} \\ 
\hline
I'm having issues with my landlord not following our lease agreement. The lease says that he is responsible for maintaining the property, but there have been so many problems, like a leaking roof and broken heating, that he hasn't fixed. I’ve tried to resolve this with him, but nothing has changed. I need legal advice on how to enforce the terms of my lease. & \textbf{Eastern Missouri} \newline [...] g. Landlord Tenant Lawsuits Lawsuits or pre-lawsuits tenant has good claims against
the landlord, management, or landlord lawyer. \newline
i. Conditions: Tenant must have notified Landlord of conditions issues and landlord has failed to address issues after notice. [...]
Examples: \newline [...] 2. No heat or ventilation & Accept \\
\hline
\end{tabular}
\end{adjustbox}
\caption{Example of scenario, excerpt of intake rule and resulting decision}
\end{table}

To assess how well LLMs apply intake rules (RQ1) and identify error types (RQ2), we developed a dataset of 16 scenarios with correct initial responses from the LLM. Using ChatGPT, we generated scenarios based on the prompt: ``Create 10 scenarios of tenants seeking legal aid for landlord issues." We ran this prompt multiple times and each scenario was reviewed and reworded to reflect common and edge cases in legal aid. We manually coded each scenario for three legal programs with expected responses: ``question," ``accept," or ``deny." The final dataset contains 48 scenario-jurisdiction pairs, as shown in Table \ref{table_d1}. 

\subsubsection{D2 - Follow-up Response Dataset}
D1 only captures the initial response by the model. In order to get an understanding of whether the model is able to accurately elicit information from users (RQ3) and obtain a better qualitative understanding of the quality of the interactions, 
we manually generated 11 additional diverse scenarios and simulated a tenant giving out just small bits of information at a time to better capture how well the LLM did at generating helpful follow-up questions. We generated transcripts for these 11 scenarios. 

\subsection{Models}
We tested 8 well-known large language models, including 2 open source models (\texttt{Llama 3.1 70b Instruct} \& \texttt{405B Instruct}) and 6 commercial models from Anthropic (\texttt{claude-3-5-sonnet-20240620}), Google (\texttt{Gemini 1.5 Pro}), and OpenAI (\texttt{gpt-4o-2024-08-06}, \texttt{gpt-4-0613}, \texttt{gpt-4-turbo-2024-04-09}, \texttt{gpt-4o-mini-2024-07-18}). We chose these models to represent a wide variety of popular commercial models at different price points and open-source models, which could be important for legal aid providers due to confidentiality concerns.

\subsection{Prompt design}
\label{sec:prompt_design}
The eligibility screening prompt was designed to save time for tenants and legal aid staff, with the main goal of avoiding false negatives (i.e., ensuring qualified applicants aren't mistakenly rejected). It was developed iteratively through tests in both the ChatGPT interface and embedded in the Docassemble application. To prevent overfitting, the iteration phase did not include any scenarios from dataset D1.

Initially, the model would give inappropriate advice (e.g., suggesting the tenant work things out with their landlord), which we addressed by clarifying that its task was to determine whether the user met the minimum intake criteria, not provide legal advice. We also found that giving examples led to hallucinations, so the prompt omits example replies. The full code and prompt are available on GitHub in two repositories. \footnote{: \url{https://github.com/lemmaLegalConsulting/docassemble-MOHUDEvictionProject/} and \url{https://github.com/SuffolkLITLab/docassemble-ALToolbox}.}

\subsection{Experiments}
We evaluated LLMs’ ability to apply intake rules through 3 experiments.

\subsubsection{E1 - Prediction of the correct initial response}
To examine how well LLMs are able to apply the intake rules to new scenarios (RQ1), we set up a pipeline that would assemble a prompt based on our developed instructions (see section \ref{sec:prompt_design}) as ``system message'', together with the intake rules from one of the jurisdictions (see section \ref{sec:intake_rules}) and one of the scenarios (D1, see section \ref{sec:scenarios}) as ``user message." This pipeline was then run for all of the scenario-jurisdiction pairs (48), for each of the selected models (8), yielding a total of 384 results. For each of the results, we captured the prediction (i.e. accept, deny, question) as well as the narrative explanation provided by the LLM. The dataset and code to reproduce this experiment is available online.\footnote{URL will be added to camera-ready paper.}

\subsubsection{E2 - Analysis of LLM errors}
In order to understand the types of errors (RQ2), we manually analyzed the instances where our best-performing model (in our case, GPT-4-turbo) had a prediction that disagreed with our manually assigned results. We investigated both the type of error and the LLM-provided explanation.

\subsubsection{E3 - Qualitative Analysis of full transcripts}
To understand the quality of overall conversations (RQ3),
we had one expert rater\footnote{Terry Lawson, Managing Attorney.} employed by Legal Services of Eastern Missouri review the 11 transcripts (D2) of manually created conversations with the Docassemble application that used GPT-4-turbo. We asked the rater to assess the transcripts' accuracy and overall quality. The rater was also asked to provide comments on each transcript.

\section{Results}

\subsection{E1 - Prediction of correct initial response}
Table \ref{tab:E1_all_models} shows the prediction performance for all models on the scenario-jurisdiction pairs, for the possible answers (Accept, Deny, Question) and the Overall average (weighted average, by proportion of class). Overall, models have a weighted F1-score of between 0.56 and 0.82, with the best model being GPT-4-Turbo. We can further see that precision for the   ``Deny'' class is very high, which we will come back to in the discussion.

Table \ref{tab:E1_per_jurisdiction} shows the performance of the best-performing model (GPT-4-Turbo) on the various jurisdiction intake rules. We can see that the model struggles the most with the Mid-Missouri intake rules, while the Eastern Missouri rules seem the easiest to apply, with a weighted average F1-score of 0.94.

\subsection{E2 - Error Analysis}
\begin{figure}[h!]
    \centering
    \begin{subfigure}[b]{0.45\textwidth}
        \centering
        \includegraphics[width=\textwidth]{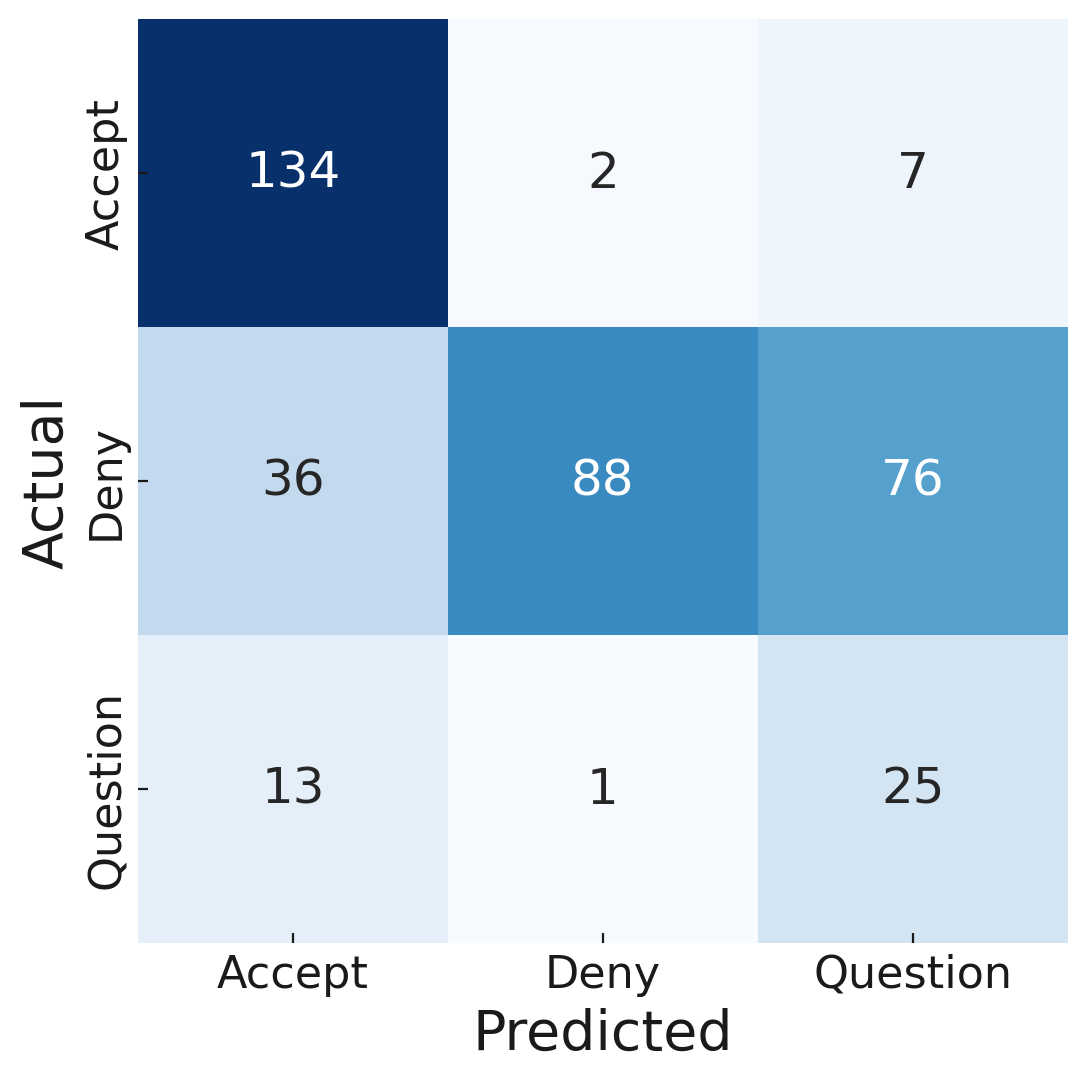}
        \caption{Confusion Matrix for All Models}
        \label{fig:all_models}
    \end{subfigure}
    \hfill
    \begin{subfigure}[b]{0.45\textwidth}
        \centering
        \includegraphics[width=\textwidth]{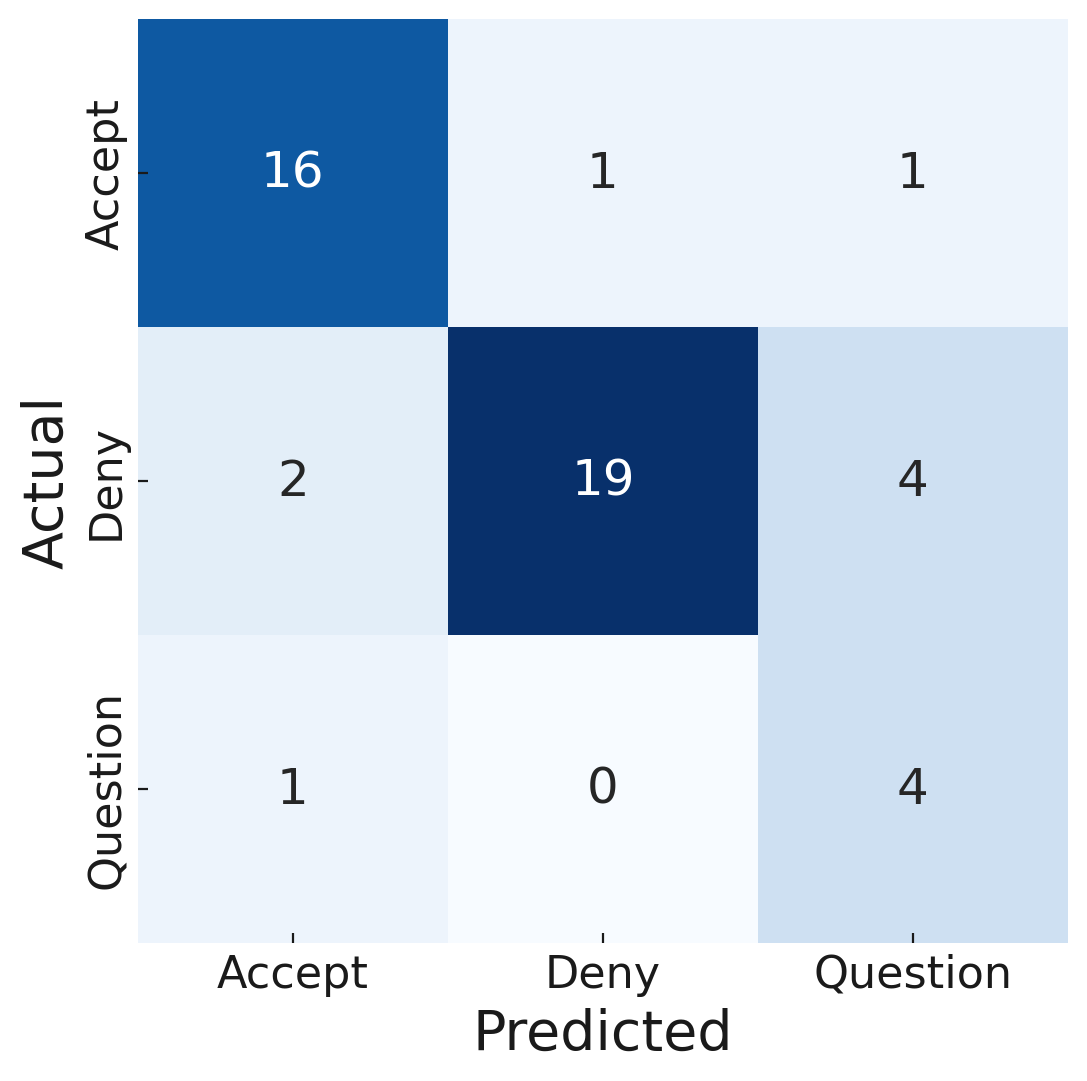}
        \caption{Confusion Matrix for gpt-4-turbo}
        \label{fig:gpt4_turbo}
    \end{subfigure}
    \caption{Confusion Matrices Comparison}
\end{figure}

We analyzed the errors made by the model, both quantitatively and qualitatively. Figure \ref{fig:all_models} shows a confusion matrix for all models, while figure \ref{fig:gpt4_turbo} shows the confusion matrix for GPT-4-Turbo, our best-performing model. We can see that the most common error is for the model to predict a situation that we classified as ``Deny'' as ``Question''. This occured 76 times for all models, and 4 times for GPT-4-turbo. It is very rare for a model to misclassify an accept or question scenario as deny.

We further analyzed these LLM errors and found 2 cases where the LLM revealed annotation mistakes, 1 case of ambiguous rules, and 6 actual LLM errors. We examine these results in-depth in \ref{sec:discussion_rq2}.


\begin{table}[]
\centering
\caption{Precision (P), Recall (R) and F1-score (F1) for the Accept, Deny and Question, and Overall (Weighted average) classes, by model. Bold shows the best score for a certain column.}
\label{tab:E1_all_models}
\begin{tabular}{|l|ccc|ccc|ccc|ccc|}

\hline
{} & \multicolumn{3}{c|}{Accept} & \multicolumn{3}{c|}{Deny} & \multicolumn{3}{c|}{Question} & \multicolumn{3}{c|}{Weighted average} \\
Model & P & R & F1 & P & R & F1 & P & R & F1 & P & R & F1 \\
\hline
Claude 3.5 Sonnet & 0.71 & 0.94 & 0.81 & \textbf{1.00} & 0.28 & 0.44 & 0.18 & 0.60 & 0.27 & 0.80 & 0.56 & 0.56 \\
Llama 3.1 405B & 0.69 & \textbf{1.00} & 0.82 & \textbf{1.00} & 0.32 & 0.48 & 0.21 & 0.60 & 0.32 & 0.80 & 0.60 & 0.59 \\
Llama 3.1 70B & 0.72 & \textbf{1.00} & 0.84 & \textbf{1.00} & 0.32 & 0.48 & 0.20 & 0.60 & 0.30 & 0.81 & 0.60 & 0.60 \\
Gemini 1.5 Pro & 0.79 & 0.83 & 0.81 & \textbf{1.00} & 0.36 & 0.53 & 0.22 & \textbf{0.80} & 0.35 & 0.84 & 0.58 & 0.62 \\
GPT-4 & 0.72 & \textbf{1.00} & 0.84 & \textbf{1.00} & 0.40 & 0.57 & 0.23 & 0.60 & 0.33 & 0.81 & 0.65 & 0.65 \\
GPT-4-turbo & \textbf{0.84} & 0.89 & 0.86 & 0.95 & \textbf{0.76} & \textbf{0.84} & \textbf{0.44} & \textbf{0.80} & \textbf{0.57} & \textbf{0.86} & \textbf{0.81} & \textbf{0.82} \\
GPT-4o & 0.81 & 0.94 & \textbf{0.87} & \textbf{1.00} & 0.44 & 0.61 & 0.19 & 0.60 & 0.29 & 0.84 & 0.65 & 0.67 \\
GPT-4o-mini & 0.62 & 0.83 & 0.71 & 0.89 & 0.64 & 0.74 & 0.33 & 0.40 & 0.36 & 0.73 & 0.69 & 0.69 \\

\hline

\end{tabular}
\end{table}

\begin{table}[]
\centering
\caption{Precision (P), Recall (R) and F1-score (F1) for the Accept, Deny and Question, and Overall (Weighted average) classes for GPT-4-Turbo, segmented by jurisdiction. Bold shows the best score for a certain column.}
\label{tab:E1_per_jurisdiction}
\begin{tabular}{|l|ccc|ccc|ccc|ccc|}
\hline
{} & \multicolumn{3}{c|}{Accept} & \multicolumn{3}{c|}{Deny} & \multicolumn{3}{c|}{Question} & \multicolumn{3}{c|}{Weighted average} \\
Jurisdiction & P & R & F1 & P & R & F1 & P & R & F1 & P & R & F1 \\
\hline
Eastern Missouri  & \textbf{1.00} & 0.88 & 0.93 & 0.88 & \textbf{1.00} & \textbf{0.93} & \textbf{1.00} & \textbf{1.00} & \textbf{1.00} & \textbf{0.95} & \textbf{0.94} & \textbf{0.94} \\
Mid-Missouri & 0.70 & 0.88 & 0.78 & \textbf{1.00} & 0.50 & 0.67 & 0.33 & 0.50 & 0.40 & 0.77 & 0.69 & 0.69 \\
Western Missouri  & \textbf{1.00} & \textbf{1.00} & \textbf{1.00} & \textbf{1.00} & 0.75 & 0.86 & 0.40 & \textbf{1.00} & 0.57 & 0.93 & 0.81 & 0.84 \\
\hline
\end{tabular}
\end{table}

\subsection{E3 - Qualitative analysis}\label{sec:qualitative_analysis}
In 73\% of cases, the expert rater marked the overall result as correct. In 63\% of cases, follow-up questions were missed that would have improved the intake. However, the human rater gave perfect scores (5 out of 5) for understandability and overall satisfaction with the tool. The human rater identified specific domain knowledge gaps, such as recognizing the severity of conditions like flimsy doors or locks.

\section{Discussion}
We investigated whether large language models can correctly determine eligibility of scenarios under complex, jurisdiction-specific intake rules. Let us explore what the results can tell us regarding our research questions.

\subsection{RQ1 - How accurately can large language models apply intake rules to legal scenarios described by laypersons?}
In E1, we compared the performance of 8 different LLMs on the task of deciding whether a scenario is eligible, not eligible, or more information is required to make this assessment. A number of interesting points are revealed by this analysis.

On a per-model basis, GPT-4-turbo performs the best, with an F1-score of .13 over the next best option. This is somewhat surprising, as we expected that more recent models such as GPT-4o and Claude 3.5 sonnet would perform better, given that they are higher on e.g. general chat leaderboards.\footnote{\url{https://lmarena.ai/}} We believe the performance discrepancy can be explained by the fact that we used GPT-4-Turbo to refine the prompt. This shows the importance of refining the prompt for specific models.

With regards to the jurisdictions, we see that there is a large performance difference between the intake rules of the various jurisdictions (Table \ref{tab:E1_per_jurisdiction}). The overall ``difficulty'' ranking between the jurisdictions is: Mid-Missouri (Avg F1-Score 0.69), Western Missouri (0.84) and Eastern Missouri (0.94). The form of the intake rules may contribute here - Eastern Missouri has the longest and most detailed rules, while Mid-Missouri features a long list of cases that are covered, which may have confused the model.


Overall, we believe that the results are strong, showing that LLMs are able to identify relevant intake rules from a set of many complex rules and apply these rules to often ambiguous scenarios with high performance, in a zero-shot setting. Next, let us take a look at the types of errors made by the LLMs.


\subsection{RQ2 - What types of errors do LLMs make, and how do they compare to human intake staff?}
\label{sec:discussion_rq2}
To better understand the results of the analysis, we quantitatively and qualitatively explored the type of errors made by the model.

The most common error was unnecessary follow-up questions (83 cases), while incorrect denials (3) and incorrect acceptances (49) were less frequent (see Figure \ref{fig:all_models}). This indicates that even the worst-performing models could be seen as safe for screening, as they prefer clarifying details or accepting cases, rather than rejecting cases.  

We paid close attention to the errors made by our best performing model, GPT-4-turbo. In two cases, the LLM helped us spot an error in our own reasoning, namely the fact that for one city in Western Missouri, ``risk of eviction'' is sufficient for acceptance. Our human error appears to have artificially lowered the performance of the GPT-4-turbo model and shows the potential for large language models to help even experts in the domain to more closely read and understand similar rules (c.f. \cite{janatian_text_2023}). In one case, we rated the answer as ambiguous - the rules themselves did not contain enough information to ascertain that risk of eviction does not encompass post-eviction cases. In 5 cases, the LLM made an understanding mistake, for example failing to grasp that ``facing eviction'' requires some imminence, failing to spot a requirement that security deposit claims require a value of over \$300 USD to qualify in Mid-Missouri, failing to make the link between having a support animal and disability discrimination, or failing to catch hints (``I need help fighting this in court'') that a trial date may have been set. Overall, these errors are relatively rare, and may be addressed in the future by further clarifications and structuring of the rules. Further, the models always seem to err on the side of caution.

\begin{table}[]
    \centering
    \begin{tabular}{|c|p{0.8\textwidth}|}
         \hline
         \textbf{Category} & \textbf{Details} \\
         \hline
         \textbf{Question} & I have a support animal, but my landlord says I cannot have it and wrote me an angry letter threatening to throw me out. \\
         \textbf{LLM response} & What city do you live in? Also, has your landlord filed an eviction against you in court? \\
         \textbf{Target} & Deny \\
         \textbf{Explanation} & This was a human error. The LLM correctly caught that in Kansas City, tenants \textbf{at risk} of eviction, even if pre-court, qualify for assistance.\\
         \hline
    \end{tabular}
    \caption{Analysis of selected errors by GPT-4-turbo in the program serving Kansas City and surrounding suburbs}
    \label{tab:error_analysis}
\end{table}


\subsection{RQ3 - Can LLMs effectively elicit missing information through follow-up questions?} 
The expert attorney's review of the dataset in D2 is encouraging. In many cases, the finding was that additional questions could have been asked by the AI model, but overall, our human gave the system perfect marks for satisfaction. This supports our belief that the system can be a useful tool for applicants to legal aid. We suspect that the goal of our expert rater, who is a supervising attorney, to obtain detailed information from the applicant to use in making a final representation decision, may not have perfectly matched the design of our system, which was to assess the minimum standard for eligibility, in order to get an applicant who may qualify to a human for a full intake as efficiently as possible.

\subsection{Limitations}
While our results are promising, we recognize that they may also have limitations.
\begin{enumerate}
    \item \textbf{Content censorship may limit applicability to other legal topics}, including topics involving violence, abuse, or unethical behavior. Google Gemini censored a scenario that stated ``I'm a DV survivor, and my landlord won’t renew my lease because they called the cops on my boyfriend."
    \item \textbf{Our prompt was iteratively refined on GPT-4-turbo}, but adjusting our prompt could improve performance on other generally high performing models.\cite{chiang2024chatbot}
    \item \textbf{Biased LLM training data may expose applicants to risks}. It is known that LLMs can reproduce biased outputs, which may be a special concern with the vulnerable legal aid population.\cite{bender_stochastic_parrot, crawford_there_2016, engstrom_algorithmic_accountability} We mitigate this risk by keeping a human in the loop, focusing the LLM on only the \textbf{minimum qualification} criteria, as well as prompting the LLM to explain the basis for its recommendation.
\end{enumerate}

\section{Conclusion and future work}
Our results show that large language models can open doors in the legal intake process, reducing time spent by both applicants and staff.  The errors observed were minor, similar to those a human intake worker might make, and sometimes even helped us catch our own mistakes. A qualitative review showed the models asked thoughtful questions and made sound recommendations, while preferring clarification or accepting the scenarios in case of doubt, a highly desirable property for pre-screening tools. 

Though our approach is promising, legal intake is a rich area for further improvements. Next steps include integrating the tool with a seamless online intake experience, improving user analytics, and simplifying rule updates by allowing staff to upload documents directly. We also see potential in using semi-structured reasoning (c.f. \cite{westermann2024dallma}) and refining prompts and intake rules to improve performance. Additionally, we plan to evaluate human intake staff performance to better understand how LLMs compare and to further explore potential biases in the LLM's recommendations. Finally, we see potential for this system to help with best match eligibility recommendations across multiple providers rather than single-provider eligibility, further expanding access to the appropriate help.

By improving eligibility information and reducing barriers, LLMs can help more people find the legal assistance they need, representing a crucial step toward a more accessible and efficient legal system.

\bibliographystyle{acm}
\bibliography{bibliography}

\end{document}